# Pressure-enhanced ferromagnetism in layered $CrSiTe_3$ flakes

Cheng Zhang[+1,2], Yue Gu[+3], Le Wang[+2], Lianglong Huang[2], Ying Fu[2], Cai Liu[2], Shanmin Wang[2], Jia-Wei Mei[*2,5], Xiaolong Zou[*3], Jun-Feng Dai[*2,4]

1. School of Physics, Harbin Institute of Technology, Harbin, 150001, People's Republic of China
2. Shenzhen Institute for Quantum Science and Engineering, and Department of Physics, Southern University of Science and Technology, Shenzhen, 518055, China
3. Shenzhen Geim Graphene Center, Tsinghua- Berkeley Shenzhen Institute & Tsinghua Shenzhen International Graduate School, Tsinghua University, Shenzhen 518055, P. R. China
4. Shenzhen Key Laboratory of Quantum Science and Engineering, Shenzhen 518055, China
5. Shenzhen Key Laboratory of Advanced Quantum Functional Materials and Devices, Southern University of Science and Technology, Shenzhen 518055, China

[+] The authors contribute to this work equally
* Corresponding authors:
daijf@sustech.edu.cn; xlzou@sz.tsinghua.edu.cn; meijw@sustech.edu.cn

**Abstract:**

**The research on van der Waals (vdW) layered ferromagnets have promoted the development of nanoscale spintronics and applications. However, low-temperature ferromagnetic properties of these materials greatly hinder their applications. Here, we report pressure-enhanced ferromagnetic behaviours in layered $CrSiTe_3$ flakes revealed by high-pressure magnetic circular dichroism (MCD) measurement. At ambient pressure, $CrSiTe_3$ undergoes a paramagnetic-to-ferromagnetic phase transition at 32.8 K, with a negligible hysteresis loop, indicating a soft ferromagnetic behaviour. Under 4.6 GPa pressure, the soft ferromagnet changes into hard one, signalled by a rectangular hysteretic loop with remnant magnetization at zero field. Interestingly, with further increasing pressure, the coercive field ($H_c$) dramatically increases from 0.02 T at 4.6 GPa to 0.17 T at 7.8 GPa, and the Curie temperature ($T_c^h$: the temperature for closing the hysteresis loop) also increases from ~36 K at 4.6 GPa to ~138 K at 7.8 GPa. The influences of pressure on exchange interactions are further investigated by density functional theory calculations, which reveal that the in-plane nearest-neighbor exchange interaction and magneto-crystalline anisotropy increase simultaneously as pressure increases, leading to increased $H_c$ and $T_c^h$ in experiments. The effective interaction between magnetic couplings and external pressure offers new opportunities for both searching room-temperature layered ferromagnets and designing pressure-sensitive magnetic functional devices.**

Recently, layered ferromagnets coupled by van der Waals (vdW) force such as $CrGeTe_3$,[1] $CrI_3$,[2] and $Fe_3GeTe_2$,[3, 4] have gained significant interest, because they not only provide an ideal platform to study the fundamental physics of magnetic interaction in two-dimensional (2D) limit, but also offer great potential in the nanoscale spintronics applications. So far, the intrinsic Curie temperatures ($T_c$) of known 2D ferromagnets are much lower than room temperature ($CrI_3$, 45K; $GrGeTe_3$, 30 K; $Fe_3GeTe_2$, 68 K),[1-3] hence, considerable efforts have been devoted to increasing their $T_c$ and making the long-range ferromagnetism more robust. Naturally, enhancing the ferromagnetic exchange interaction can contribute to the increase of $T_c$.[5] For 2D ferromagnets, magneto-crystalline anisotropy breaks the spin-rotational symmetry and can also stabilize 2D Ising-type ferromagnetism. Therefore, searching for new routes to improve these two magnetic interactions is of great significance to the applications of 2D ferromagnets.

Compared with bulk cases, the magnetic states of 2D ferromagnets show greater controllability by external stimuli,[3, 6-11] which may improve their magneto-crystalline anisotropy and exchange interaction, thereby expanding their applications and deepening our understanding of the underlying mechanism of magnetism in 2D systems. Hydrostatic pressure is one of the practical and easy-to-implement routes for controlling 2D magnetic states[12-16]. It can effectively tune the distance between adjacent atomic layers, and bond lengths and bond angles of intralayer atoms in 2D magnets, leading to modulation of interlayer and intralayer exchange interactions as well as magneto-crystalline anisotropy. Experimentally, for bilayer $CrI_3$,[7, 17] hydrostatic pressure can switch the magnetic orders between antiferromagnetic and ferromagnetic states. Under pressure, $Fe_3GeTe_2$ shows the reduction of the local moment on Fe ions as well as pressured-induced increase of the electronic itinerancy.[18, 19] Moreover, a spin reorientation is reported in layered ferromagnetic insulator $CrGeTe_3$ under pressure,[20] highlighting the effective control of magneto-crystalline anisotropy by pressure.

With similar crystal structure as $CrGeTe_3$ (Figure 1a), $CrSiTe_3$ also exhibits soft ferromagnetic behavior without any remnant magnetization under zero field.[21] After the replacement of Ge by smaller Si, the application of hydrostatic pressure should further shorten the distance between $3d$ atoms. Therefore, the exchange interaction and the magnetic behaviors could be greatly influenced given the small bandgap of $CrSiTe_3$.[22] Meanwhile, it has been reported that ferromagnetic $CrSiTe_3$ undergoes a transition to superconducting phase under a pressure of ~7.5 GPa,[23] which makes a detailed understanding of the evolution of magnetic behaviors very important. In this work, a $CrSiTe_3$ flake was capsulated in a diamond anvil cell (DAC) with pressure up to 10 GPa and the ferromagnetic behaviours under pressure were investigated via in-situ high-pressure magnetic circular dichroism (MCD) microscopy. We found that $CrSiTe_3$ experiences a transition from soft to hard ferromagnetic states with a remanent magnetization at zero field appearing beyond 4.6 GPa. Importantly, a further increase of pressure up to 7.8 GPa dramatically increases the coercive field ($H_c$) and Curie temperature ($T_c^h$), indicating the enhancement of magneto-crystalline anisotropy and exchange interaction. Our density functional theory (DFT) calculations also reveal enhanced magneto-crystalline anisotropy energy and in-plane nearest-neighbour exchange interaction ($J_1$) as a function of pressure, corroborating experiment observations.

High pressure experiments were performed in a DAC with pressure up to 10 GPa. In order to load sample into DAC, we first mechanically exfoliated CrSiTe$_3$ flakes from signal crystal with a size larger than $50 \times 50\ \mu m$. However, the sample thickness was in the scale of nanometer (~100 nm). Then they were transferred onto the culet of diamond using PC films. The entire sample was then covered by BN flakes to prevent sample degradation. After filling the pressure-transmitting medium and ruby balls, the entire DAC was mounted onto the cold-finger of low-temperature cryostat with temperature in the range between 16 K and 300 K. The out-of-plane ferromagnetic states were probed using in situ high-pressure MCD spectroscopy. Figure 1b shows a schematic diagram of experimental setup for MCD measurement in a reflection geometry under high-pressure environment. The details about this technique can be found in method section.

High quality CrSiTe$_3$ single crystals were grown by the flux method (see method for details). The representative sample size was around $5 \times 7\ mm^2$ as shown in the inset of Figure 1c. CrSiTe$_3$ has a 2D honeycomb layered structure coupled by vdW force (Figure 1a). The XRD results in Figure 1c and Figure S1 indicate the high quality of our sample. And the magnetic susceptibility ($\chi$) measurement in Figure 1d reveals a ferromagnetic order (FM) with Curie temperature $T_c$ of 32.8 K. The value of $\chi_c$ is more than four times $\chi_{ab}$ at 2 K, which suggests that the crystallographic c-axis is the easy axis, consistent with the previous report.[24] The MCD measurement also shows a saturated magnetization as the applied field increase above 0.05 T at 16 K (Figure 1e) due to the ferromagnetic properties. However, a hysteresis loop is absent in MCD measurement when we sweep the field back and forth. It indicates a soft ferromagnetic behavior in bulk CrSiTe$_3$, which has a similar magnetic behavior as that of CrGeTe$_3$[9]. Overall, at ambient pressure, CrSiTe$_3$ exhibits a quasi-2D Ising ferromagnetic orders below a Curie temperature of ~32.8 K with a negligible coercivity.

The pressure-induced MCD changes will reveal more important information about interlayer or intralayer exchange interaction due to the shrink of distance between magnetic atoms. To study the evolution of magnetic properties of CrSiTe$_3$, we measure the MCD signals in a CrSiTe$_3$ flake at a temperature of 16 K under five representative hydrostatic pressures of 0.7, 2.5, 4.6, 6.2, and 7.8 GPa, respectively. As shown in Figure 2a, at 0.7 and 2.5 GPa, the magnetic behaviour of CrSiTe3 is characterized by a linear relation between the magnetization and the applied magnetic field below ±0.15 T and a saturated magnetization above that value. However, no measurable hysteresis loop, namely, remanent magnetization at $\mu_0 H = 0\ T$, can be found under two pressures. Although the saturated magnetization indicate ferromagnetic behaviour due to spin-spin exchange interaction, no hysteresis loop suggests that the barrier between two spin polarized states is very low. So this is considered as the soft ferromagnetic state. Surprisingly, at 4.6 GPa, a typical rectangular hysteresis loop appears as applied field scans in the same range. The average coercive field is evaluated to be 0.02 T at 16 K. Because of the appearance of remnant magnetic order at zero field, we consider the corresponding magnetic state of CrSiTe$_3$ as the hard ferromagnetic state. As pressure increases further, the area of hysteresis loop increases and the corresponding coercive field also increases from 0.02 T at 4.6 GPa to 0.17 T at 7.8 GPa. Above 7.8 GPa, we cannot detect any MCD signal due to a structural transition at ~ 7.5 GPa accompanied by a ferromagnetic-paramagnetic phase transition[11] (data are not shown). As summarized in Figure 2b, the coercive field dramatically increases as pressure increases, with a threshold pressure around 4 GPa. Since the coercive field $H_c$ is proportional to the easy-axis single-ion magnetic anisotropy, we infer that the uniaxial anisotropy energy in CrSiTe$_3$ dramatically increases above threshold pressure, which is significantly different from the magnetic behaviors of Fe$_3$GeTe$_2$ under pressure[7].

Figure 3 shows the temperature-dependence of MCD measurement as a function of magnetic field under different pressures. At 0.7 and 2.5 GPa, the MCD signals gradually change from a linear relation with saturation to a pure linear relation within ±0.3 T as the temperature increases. This indicates a transition from ferromagnetic to paramagnetic states, consistent with susceptibility measurement at normal pressure. From the temperature dependence of MCD signals, we can roughly extract that the transition temperature at two pressures is ~ 32 K at 0.7 GPa (Figure 3a) and ~ 34 K at 2.5 GPa (Figure 3b), respectively. Compared with the Curie temperature at normal pressure, the negligible difference of transition temperature indicates the weak influence of pressure on magnetic behaviour.

According to mean field theory for ferromagnetic Ising model, the critical transition temperature is proportional to spin-spin exchange interaction (J). Obviously, J is not dramatically changed in this pressure region. For the hard ferromagnetic state at 4.6 GPa (Figure 3c), the hysteretic loop can remain until 40 K. Here, we define the temperature for the closure of hysteresis loop as the Curie temperature for hard ferromagnetic state ($T_c^h$) and mainly discuss the influence of pressure on it. As pressure increases further, the hysteresis loop can survive up to higher temperature, which is evaluated to be 100 K at 6.2 GPa (Figure 3d) and 138 K at 7.8 GPa (Figure 3e), respectively. Since the proportional relation between Curie temperature and J in Ising model, the dramatically increases of $T_c^h$ indicates the inter- or intra-layer exchange interaction is enhanced by passing through a threshold pressure (around 4 GPa). Experimentally, the increased pressure mainly decreases the interlayer atomic distance and intralayer bond length and angle between atoms, which correspond to tune the interlayer and intralayer exchange interaction. However, we cannot separate them simply based on our experimental measurements.

To understand the origin of increasing $T_c$ and $H_c$ under pressure, DFT calculations were employed to investigate the response of structural, electronic, and magnetic properties of CrSiTe$_3$ to external pressure. CrSiTe$_3$ crystallizes in the $R\bar{3}$ space group, with the monolayers stacked in ABCABC sequence, as shown in Figure 1a. To describe possible structural transition, we also consider two other stacking orders, i.e., AA and inclined AA sequences illustrated in the insets of Figure S2, both of which correspond to $C2/m$ space group. The energies of these three structures as a function of pressure is shown in Figure S2. It is noted that even when the pressure increases to 12 GPa, the energies of these two $C2/m$ phases are still higher than that of $R\bar{3}$ phase. Therefore, we focus on the $R\bar{3}$ phase hereafter.

To explore the magnetic properties of CrSiTe$_3$ under pressure, we have considered various magnetic orders, including ferromagnetic (FM), interlayer antiferromagnetic (AFM) with intralayer FM coupling, and three intralayer AFM with interlayer FM coupling. The four AFM states are named as interlayer-AFM (iAFM), Neel-AFM (nAFM), Stripy-AFM (sAFM), and Zigzag-AFM (zAFM). The last three states are shown in Figure 4a, according to their different intralayer arrangement for local magnetic moments. The calculated pressure-dependent energies for these states by DFT+U method are displayed in Figure 4b, which clearly shows that FM is always the ground state. More importantly, the energy difference between FM and AFM orders generally increases as pressure increases, which suggests that the stability of ferromagnetism is enhanced by pressure. The evolution of band structures under pressure is shown in Figure S3, along with the orbital projection to Cr and Te atoms. The top valence bands and bottom conduction bands are contributed by Te atoms and hybrid Cr and Te orbitals, respectively. Under pressure, top valence bands shift up significantly, whereas bottom conduction bands move down slightly. Accordingly, the metallicity mainly originates from Te orbitals, and the Mott and magnetic behaviors remain largely unchanged.

To give a quantitative estimation of $T_c$ of CrSiTe$_3$, we adopted mean-field theory (MFT). Based on the Heisenberg Hamiltonian $H = \sum_{i,j} J_{ij} \mathbf{S}_i \cdot \mathbf{S}_j$, where $S_i$ is the spin operator of Cr on site i (~ 3/2 in our case), and $J_{ij}$ is the exchange coupling constant between spins on site i and j, $T_c$ can be evaluated as $T_c = -\frac{2S(S+1)}{3k_B}\left(\frac{1}{n}\sum_{i \neq j} J_{ij}\right)$ with $n$ the number of neighboring sites.[3] Here, the intralayer first, second and third nearest-neighbor exchange interactions $J_1$, $J_2$, and $J_3$, as well as interlayer first nearest-neighbor exchange interaction $J_4$ were considered, as illustrated in the insets of Figure 4c. As shown in Figure 4c, $J_2$ having the smallest absolute value varies in a narrow range between -0.5 and 0.5 meV. In contrast, $J_1$, $J_3$ and $J_4$ are always negative, which tend to stablize FM order. Under pressure, $J_3$ shows a slow and monotonic decline, while $J_1$ and $J_4$ decrease rapidly after 4 GPa, with a slight increase for $J_4$ after 8 GPa. In particular, the absolute values of $J_1$ and $J_4$ are much larger than those of $J_2$ and $J_3$, with $J_1$ always stronger than $J_4$. Given the number of neighboring sites for $J_1$ ($n$ = 3) is larger than that for $J_4$ ($n$ = 1), the contribution of $J_1$ in the Hamiltonian is much greater than $J_4$. Therefore, the interlayer first nearest-neighbor exchange interaction $J_1$ plays a dominant role in the enhancement of ferromagnetism. Except for the intial slight decrease by about 19.4 K below 3 GPa, $T_c$ increases rapidly, reaching 325K under 8 GPa, which is about 132 K higher than that under 0 GPa. Although the MFT method generally overestimates $T_c$,[25] the obtained variation of $T_c$ is consistent with our experimental results from 0 to 7.8 Gpa. The increase of FM interaction under pressure can be qualitatively understood by the change in the gap between spin-up unoccupied $e_g$ and occupied $t_{2g}$ orbitals (see schematic in Figure 4d for virtual hopping), which shows a monotonic decrease versus pressure (Figure S4), giving rise to enhanced FM exchange coupling.[26] Besides DFT+U method, we also adopted the HSE06 functional to estimate $T_c$ under different pressures by considering intralayer $J_1$, $J_2$, $J_3$ only to reduce the computational cost, and the results shows similar trend for $T_c$, as shown in Figure S5.

The experimentally observed change in $H_c$ can be rationalized by analysing magneto-crystalline anisotropy energy (MAE), which is defined as the energy difference between states with magnetic moments along in-plane and out-of-plane directions, respectively. The positive MAE indicates easy-axis magnetization, while negative one suggests easy-plane magnetization. Our calculations shown that MAE is highly sensitive to the Hubbard U parameter as shown in Table S1. Thus, we adopted the HSE06 method instead to calculated MAE, as shown in Figure 4e. It can be clearly seen that as the pressure increases, MAE keeps positive and increases monotonically from 0.39 meV/f.u. at 0 GPa to 0.46 meV/f.u. at 6 GPa, followed by a slight decrease. These results are consistent with the enhanced coercive field and harder ferromagnetic behaviors under pressure in our experiment.

In conclusion, we have demonstrated a behaviour of soft ferromagnetic to hard ferromagnetic transition in CrSiTe$_3$ flakes by means of hydrostatic pressure. The rectangular hysteretic loop, signed by a remnant magnetization at zero field, appears above 4.6 GPa, with increased coercive field ($H_c$) and Curie temperature ($T_c^h$) as pressure increases further. Our DFT calculations justified that the in-plane nearest-neighbor exchange interaction and magneto-crystalline anisotropy contribute to increased $H_c$ and $T_c^h$ in experiments. Our results will be helpful in understanding the underlying mechanism of magnetic exchange interaction and open up an novel route to effectively tune magnetic properties in 2D materials.

**Acknowledgement**

We would like to thank Prof. Haizhou Lu from SUSTech for helpful discussions. J.F. acknowledges the support from the National Natural Science Foundation of China (11974159) and the Guangdong Natural Science Foundation (2021A1515012316). J.W.M was partially supported by the program for Guangdong Introducing Innovative and Entrepreneurial Teams (No. 2017ZT07C062), and Shenzhen Key Laboratory of Advanced Quantum Functional Materials and Devices (No. ZDSYS20190902092905285). J.W.M and L.W. was supported by Guangdong Basic and Applied Basic Research Foundation (No. 2020B1515120100). L.W. was supported by China Postdoctoral Science Foundation (2020M682780). Theoretical part was supported by the National Natural Science Foundation of China (11974197, 51920105002), Guangdong Innovative and Entrepreneurial Research Team Program (No. 2017ZT07C341).

**Conflict of Interest**

The authors declare no conflict of interest.

**Keywords**

2D ferromagnetic materials, exchange interaction, high-pressure magnetic circular dichroism (MCD) spectroscopy

**Methods**

**Sample growth**

CrSiTe$_3$ crystals were grown by using the Si-Te eutectic as flux. High-purity elements Cr grains (99.996%), Si pieces (99.9999%), Te blocks (99.9999%) were weighed in the molar ratio Cr:Si:Te = 1:2:6, and placed in an alumina crucible, then sealed in a fully evacuated quartz tube. The crucible was heated to 1373 K and dwell for 10 hrs, then cooled slowly to 973 K in 150 hrs, where the flux was spun off by a centrifuge. The single crystal XRD and powder (obtained by grounding the single crystals) XRD were performed on a Rigaku Smartlab-9kW diffractometer with Cu Kα radiation ($\lambda_{K\alpha1}$ = 1.54056 Å) at room temperature. The magnetic susceptibility ($\chi$) was performed on a Quantum Design MPMS3 SQUID magnetometer.

**Magnetic circular dichroism measurements under high pressure**

The ferromagnetic behaviours of CrSiTe$_3$ flake under high pressure were measured in in situ magnetic circular dichroism (MCD) system (Figure 1b). Here, a 632.8 nm HeNe laser was selected as an excited light, and focused onto sample within DAV by a long working distance objective with spot size of around 2 μm. The polarized state of excited beam was modulated by a photoelastic modulator (PEM), so that it changed between left-handed and right-handed circularly polarization with a frequency of 50 KHz. The reflected MCD signal, namely the intensity difference between two circularly polarized lights, is collected by a Si detector and recorded by an lock-in amplifier. During measurements, the out-of-plane magnetic field, generated from a supper-conducting loop, was scanned within 0.3 T. Hence, we can get the MCD intensity as a function of applied magnetic field.

**Density functional theory (DFT) first-principles calculations**

Our density functional theory (DFT) calculations were done adopting the Vienna ab initio Simulation Package (VASP),[27, 28] with Perdew–Burke–Ernzerhof parameterization of the generalized gradient approximation (GGA)[29] and projector augmented wave (PAW) method.[30] describing the exchange-correlation and ion-electron interactions, respectively. A cutoff energy of 500 eV was chosen for plane waves basis set. The structural optimization was performed until the force on each atom and energy were converged to 0.01 eV/Å and 1 ×10$^{-6}$ eV, respectively. Van der Waals (vdW) interaction was included using the optB86b-vdW functional.[31] The DFT+U method was used to treat the onsite coulomb interaction of Cr $d$ orbitals properly.[32] Similar as CrGeTe$_3$,[1] the choice of Hubbard U is crucial to determine the electronic and magnetic properties of CrSiTe$_3$. We tested different values of U (Table S1), and finally chose U = 0.5 to ensure the appropriate ground state of CrSiTe$_3$, which shows the correct ground magnetic structure, i.e., interlayer ferromagnetic coupling and the easy-axis magnetization along the out-of-plane direction. With 6 × 6 × 2 k-point sampling, the optimized bulk lattice constants are $a$ = $b$ = 6.81 Å, and $c$ = 20.48 Å. In addition, Hybrid functional (HSE06) with 3 × 3 × 1 k-point sampling was used for the calculation of magneto-crystalline anisotropy energy.[33]

**FIGURES AND CAPTIONS**

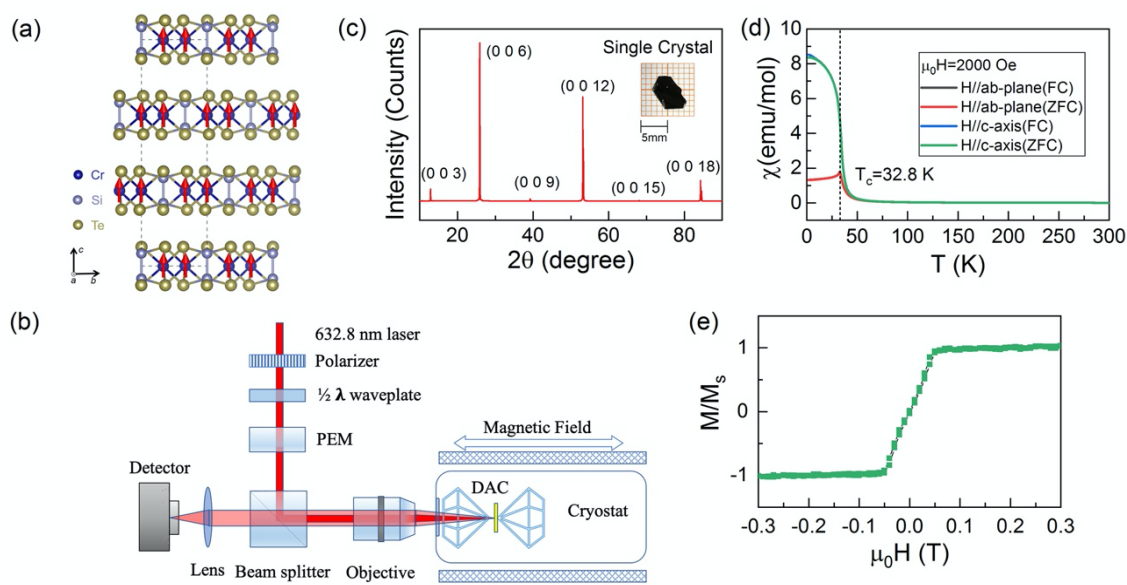

Figure 1: (a) Side view of lattice structure of CrSiTe$_3$. The red arrows indicate the magnetization direction. (b) Schematic of in situ high-pressure magnetic circular dichroism (MCD) experimental setup with the lowest temperature of 16 K and the highest pressure of 12 GPa, respectively. (c) The single crystal X-ray diffraction pattern of CrSiTe$_3$. Inset is the optical image of single crystal CrSiTe$_3$. (d) Magnetic susceptibility of CrSiTe$_3$ measured on a single crystal with field cooling (FC) and zero-field cooling (ZFC) modes under a magnetic field of 2000 Oe applied along the c-axis and ab-plane, respectively. (e) Magnetization of a CrSiTe$_3$ flake as a function of magnetic field (H) along c axis, measured at 16 K at normal pressure.

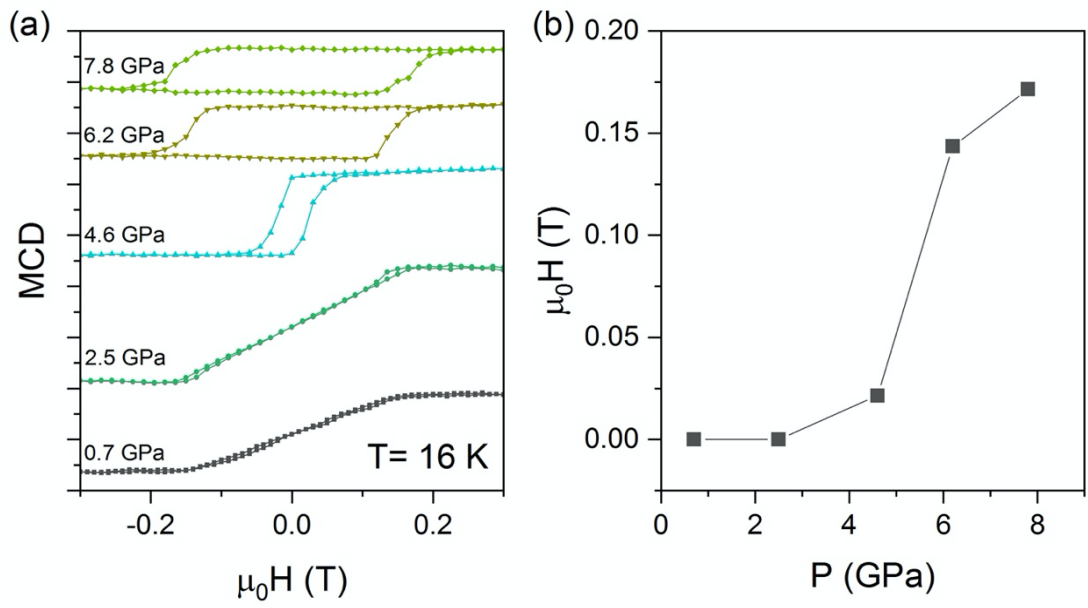

Figure 2: (a) MCD signals as a function of applied magnetic field at 16 K as pressure changes from 0.7 GPa to 7.8 GPa. (b) Extracted average critical field for spin-flip transition as a function of applied pressures.

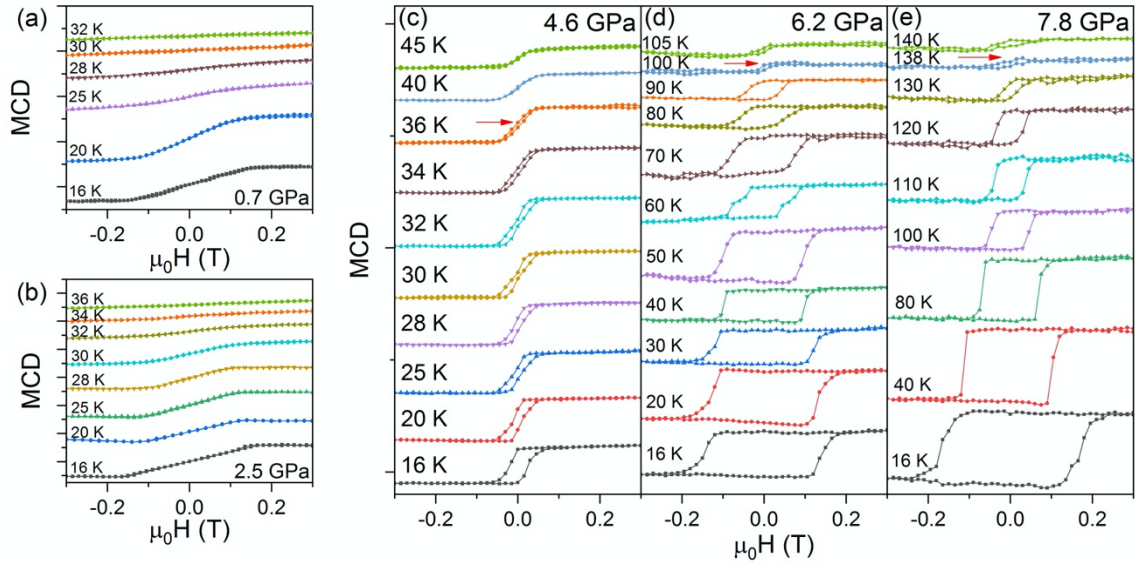

Figure 3: Temperature dependence of MCD signals in the CrSiTe$_3$ flake under five pressures, namely (a) 0.7, (b) 2.5, (c) 4.6, (d) 6.2 and (e) 7.8 GPa, respectively. The red arrows indicate the temperature for closure of hysteresis loop at 4.6, 6.2 and 7.8 GPa, respectively.

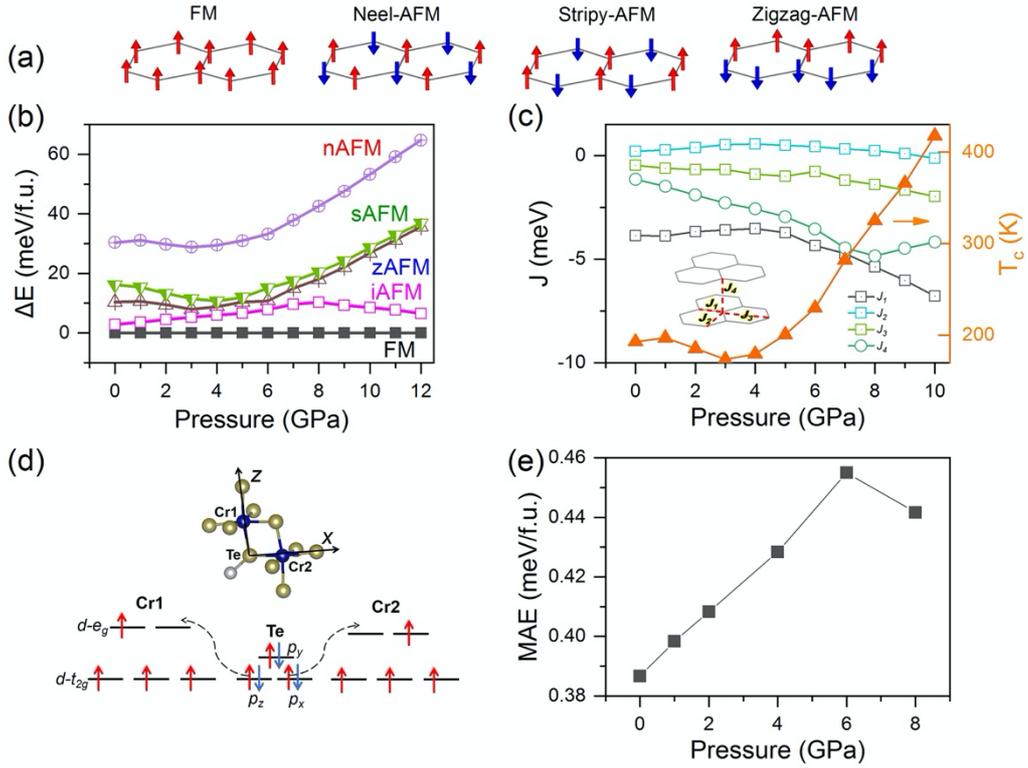

Figure 4: (a) Ferromagnetic and three antiferromagnetic orders: Neel-AFM, Stripy-AFM, and Zigzag-AFM. (b) The calculated energy difference between different magnetic orders and ground FM state as a function of pressure. (c) The calculated intralayer first, second and third nearest-neighbor exchange interactions $J_1$, $J_2$, $J_3$, interlayer nearest-neighbor exchange interactions $J_4$ (illustrated in the insets), and Curie temperature $T_c$ as functions of pressure. (d) Schematics for local coordination and virtual hopping between Cr-$d$ orbitals intermediated by Te-$p$ orbitals. (e) The calculated magneto-crystalline anisotropy energy as a function of pressure.